\documentclass[conf]{new-aiaa}
\usepackage[utf8]{inputenc}

\usepackage{graphicx}
\usepackage{amsmath}
\usepackage[version=4]{mhchem}
\usepackage{siunitx}
\usepackage{longtable,tabularx}
\usepackage{todonotes}
\usepackage{caption} % Add captions to figures
\usepackage{psfrag} % Replace text in figures
\usepackage{graphicx} % Extend graphic features (e.g. Resizebox for LaPrint)
\usepackage{float} % Place floats exactly here
\usepackage{xfrac}
\usepackage{colortbl}
\newcommand{\myTilde}{\raisebox{0.5ex}{\texttildelow}}

\title{Statistical Dependence Analyses of Operational Flight Data Used for Landing Reconstruction Enhancement}

\author{Lukas Höhndorf\footnote{Research Associate, Institute of Flight System Dynamics, lukas.hoehndorf@tum.de.}}
\affil{Technical University of Munich, Institute of Flight System Dynamics, 85748, Garching bei München, Germany}
\author{Thomas Nagler\footnote{Research Associate, Chair of Mathematical Statistics, thomas.nagler@tum.de}}
\affil{Technical University of Munich, Chair of Mathematical Statistics, 85748, Garching bei München, Germany}
\author{Phillip Koppitz\footnote{Research Associate, Institute of Flight System Dynamics, phillip.koppitz@tum.de.}}
\affil{Technical University of Munich, Institute of Flight System Dynamics, 85748, Garching bei München, Germany}
\author{Prof. Ph.D. Claudia Czado\footnote{Associate Professorship of
Applied Mathematical Statistics, Chair of Mathematical Statistics, cczado@ma.tum.de}}
\affil{Technical University of Munich, Chair of Mathematical Statistics, 85748, Garching bei München, Germany}
\author{Prof. Dr.-Ing. Florian Holzapfel\footnote{Professor and Head of the Institute of Flight System Dynamics, florian.holzapfel@tum.de.}}
\affil{Technical University of Munich, Institute of Flight System Dynamics, 85748, Garching bei München, Germany}

\begin{document}
% \begin{center}
% The authors permit this paper to be considered for the award of the 22\textsuperscript{nd} ATRS World Conference 2018 and for the ATRS Special Issue of the Journal of Air Transport Studies (JATS).
% \end{center}

\maketitle

\begin{abstract}
To analyze flight data recorded during an airline’s flight operation, a so-called Rauch-Tung-Striebel (RTS) smoother is implemented at the Institute of Flight System Dynamics. The RTS smoother is widely used for state estimation and it is utilized here to increase the data quality with respect to physical coherence and to increase resolution. The purpose of this paper is to enhance the performance of the RTS smoother to reconstruct an aircraft landing using on board recorded data only. Thereby, errors and uncertainties of operational flight data (e.g. altitude, attitude, position, speed) recorded during flights of civil aircraft are minimized. These data can be used for subsequent analyses in terms of flight safety or efficiency, which is commonly referred to as Flight Data Monitoring (FDM).

Statistical assumptions of the smoother theory are not always verified during application but (consciously or not) assumed to be fulfilled. These assumptions can hardly be verified prior to the smoother application, however, they can be verified using the results of an initial smoother iteration and modifications of specific smoother characteristics can be suggested. This project specifically verifies assumptions on the measurement noise characteristics. 

Variance and covariance of the measurement noise can be checked after the initial smoother application. It is discovered that these characteristics change over time and should be accounted for with a time varying covariance matrix. This sequence of matrices is estimated by kernel smoothing and replaces an initially assumed fixed and diagonal covariance matrix used for the first smoother run. The results of this second smoother iteration are mostly improved compared to the initial iteration, i.e.\ the errors are significantly reduced. Subsequently, the remaining dependence structures of the residuals of the second smoother iteration can be captured by copula models. Their interpretation is useful for a revision of the physical model utilized by the RTS smoother.

\vspace{0.6cm}

\end{abstract}
\newpage
\section*{Abbreviations}

{\renewcommand\arraystretch{1.0}
\noindent\begin{longtable*}{@{}l @{\quad=\quad} l@{}}
DDM & difference in the depth of modulation \\
EKF & extended Kalman filter\\
FDR & flight data recorder \\
FDM & flight data monitoring \\
GPS & global positioning system \\
ILS&instrument landing system \\
NED&north-east-down frame\\
QAR & quick access recorder \\
RALT&radio altitude\\
RTS & Rauch-Tung-Striebel smoother \\
SQM & smoothing quality measure
\end{longtable*}}

\section{Introduction}
\lettrine{D}{uring} the flight of a civil aircraft a huge amount of data is recorded. Besides the Flight Data Recorder (FDR) that is widely known as ``black box'' and is used for accident investigations there are also recorders on board the aircraft to collect data  of the regular operation. These recorders are called Quick Access Recorder (QAR) and also wireless technologies for the transmission of the recorded data get common nowadays. These data are collected, stored, and analyzed as part of the airline's Flight Data Monitoring (FDM) framework. Since the data are collected throughout the flight with a certain frequency, their data structure are time series.

The Institute of Flight System Dynamics at the Technical University of Munich (TUM) cooperates with several airlines and participates in projects to further develop the algorithms used in FDM. One of these projects is carried out with the TUM Chair of Mathematical Statistics with the goal to characterize and beneficially use statistical dependence structures of the time series recorded on board the aircraft.

Recorded data always contain errors and uncertainties. To increase robustness of any algorithm that is using operational flight data, data quality should be improved as much as possible. The most relevant recorded variables such as position, altitude, speed, acceleration, and attitude angles are all linked by physical relations. These physical relations can be captured in a physical model and used for data smoothing. To achieve this, the so-called Rauch-Tung-Striebel (RTS) smoother that is based on the Extended Kalman Filter (EKF) is used. A special focus of this paper lies on the verification of the RTS smoother assumptions, in particular regarding the measurement noise covariance matrices.

Once the RTS smoother is applied with the revised measurement noise covariance matrices, the residuals of the data can be used for a dependence analyses using copula structures. Their interpretation can be helpful for a revision of the physical model the RTS smoother is based on.

In chapter \ref{DataChapter}, we give a brief overview of the data. Subsequently, a summary of the state space models studied in this paper is given in chapter \ref{StateSpaceModel}. In chapter \ref{PhysicalModelChapter}, the main structure of the considered model and references for the full model details are discussed. The Kalman filter, Extended Kalman Filter (EKF), and the RTS smoother are briefly summarized in chapter \ref{SmootherChapter}. In the following chapter \ref{QualityMeasureChapter}, a measure to describe the quality of a RTS smoother application is described.  Subsequently, chapter \ref{NoiseCharacterization} describes a method to characterize the measurement noise dependencies effectively. Those dependencies can then be integrated into the RTS smoother in chapter \ref{IntegrationNoiseRTSChapter}. In chapter \ref{DependenceInvestigationChapter}, an interpretation of the remaining dependencies in the residuals based on copula models is given. Finally, chapter \ref{ConclusionChapter} concludes the paper. 

\section{Operational Flight Data Characteristics}\label{DataChapter}
\begin{figure}[ht]
	\centering
	\includegraphics[width=0.8\textwidth]{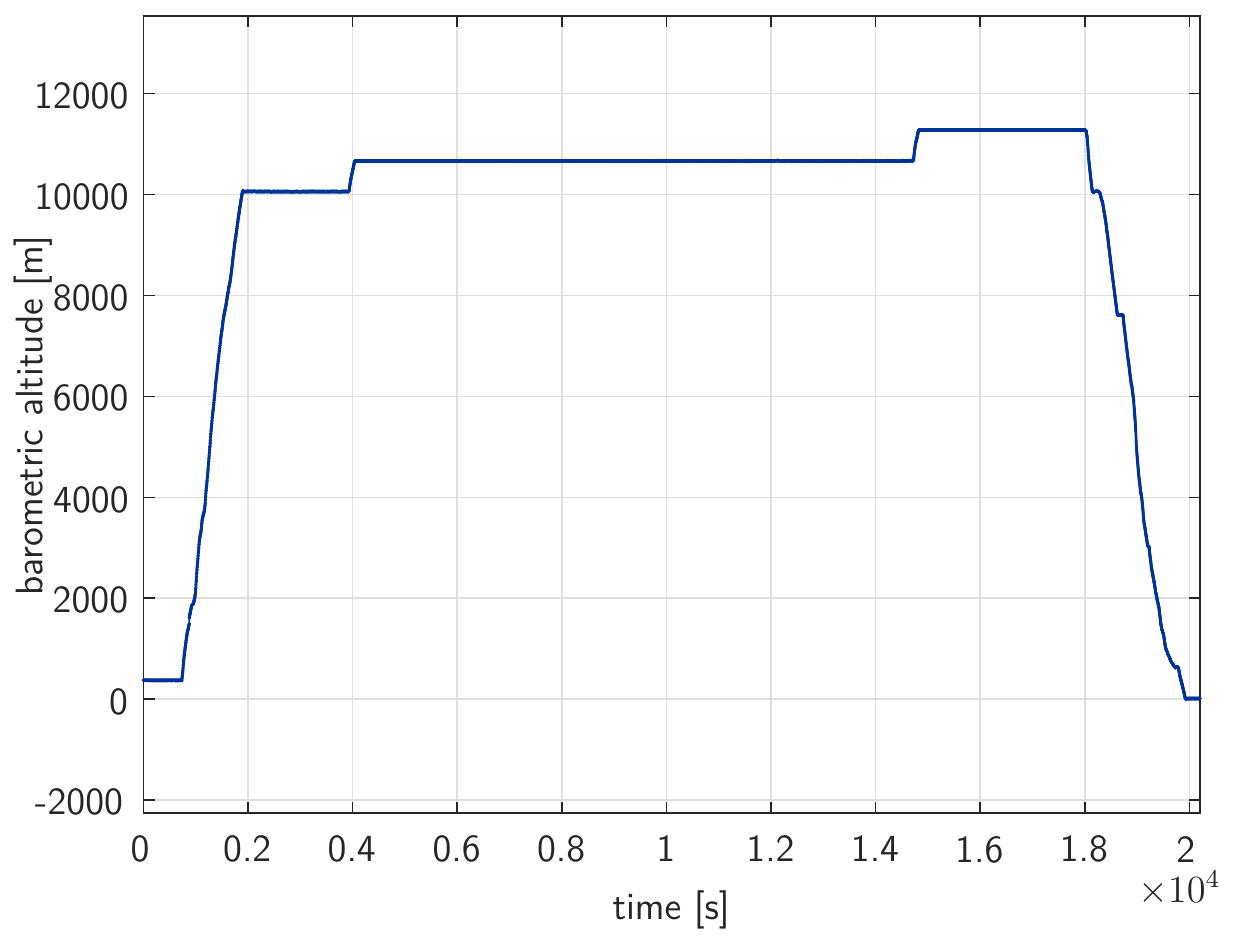}
	\caption{Time series barometric altitude}
	\label{timeSeriesBaro}
\end{figure}

Data recorded on board aircraft are collected throughout the flight and therefore as time series. The frequency of the recording commonly lies between \sfrac{1}{4} and 16 Hertz depending on the variable. In Fig.\ \ref{timeSeriesBaro} the variable barometric altitude is illustrated. Thereby the characteristic step climbs during a long haul flight to increase efficiency can be identified to take the decreasing aircraft weight due to the fuel burn into account.

For many FDM algorithms certain time points such as the touch down or lift off of the aircraft are essential. These time points can be identified based on the recorded time series. In addition, so-called measurements or snapshots can be derived. A measurement is a single value per flight that describes a particular aspect of the operation, performance, safety, or efficiency in more detail. One example is the ground speed of the aircraft at touchdown.

Identified biases and scale factors of the recorded time series that are mentioned in chapter \ref{PhysicalModelChapter} are measurements considered in this paper.

\section{State Space Model for Describing Dynamic Systems}\label{StateSpaceModel}
We represent a dynamic system by the inputs \(u\), the outputs \(y\) and the system model functions \(f\) and \(g\). These model functions link the internal system state variables \(x\) with their derivative and the outputs \(y\). The state space model is given by the state equation
\begin{equation}\label{StateEquation}
\dot{x}(t)=f(x(t),u(t),\Theta),
\end{equation}

\noindent and the output equation
\begin{equation}\label{OutputEquation}
y(t)=g(x(t),u(t),\Theta),
\end{equation}

\noindent where constant system parameters such as biases or scale factors are collected within the variable \(\Theta\).

\begin{figure}[ht]
	\centering
	\includegraphics[width=0.9\textwidth]{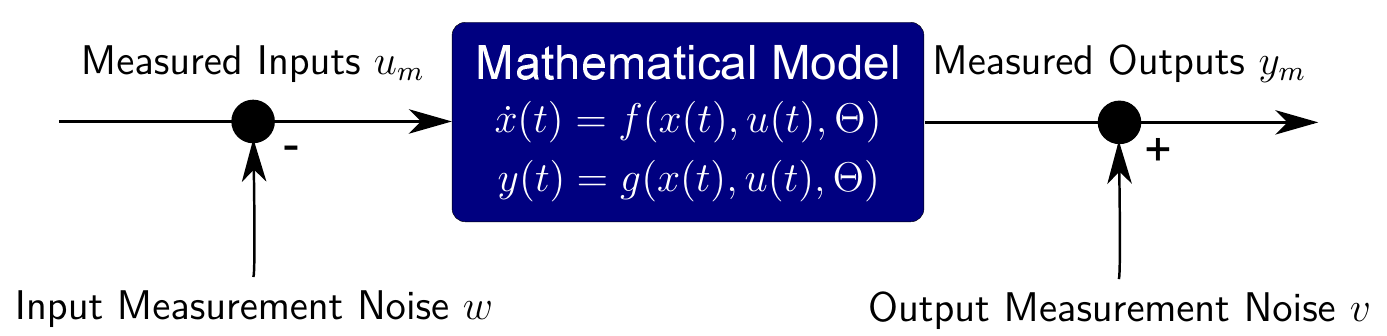}    
	\caption{Dynamic system with noise terms, based on Figure 3 of \cite{Hohndorf.2016}}
    \label{DynamicSystemOverviewWithNoise}
\end{figure}

Measurement noise is taken into account in two ways, as input measurement noise \(w\) and output measurement noise \(v\), see Fig.\ \ref{DynamicSystemOverviewWithNoise}. This leads to an adaptation of the state and output equation in the following way
\begin{gather}
	\dot{x}(t)=f(x(t),u_m(t)-w(t),\Theta)\label{StateEquationWithErrors}\\
	y_m(t)=g(x(t),u_m(t),\Theta)+v(t)\label{OutputEquationWithErrors}.
\end{gather}

\noindent Observe that the subscript \(m\) was added to the inputs \(u\) and the outputs \(y\) to highlight that measured data is considered here.

\section{Physical Aircraft Model for Landing Reconstruction}\label{PhysicalModelChapter}
The physical model that is the basis for the ideas and methods considered within this paper was mainly developed within \cite{JoachimSiegel.25.04.2017} and \cite{Siegel.18.11.2015} and also summarized in \cite{Hohndorf.2017b} and \cite{Hohndorf.2016}. It is specifically designed for application during the landing phase of the aircraft. We focus on the time period from approximately 1000 feet above ground until the aircraft enters the apron area of the airport.

Describing the complete model in all details is out of the scope of this paper. At this stage, the input, state and output vectors are given and one exemplary output equation presented. Further details can be found in the references given within this chapter.

The input vector \(u\) is given by 
\begin{equation*}\label{RTSInputVector}
\begin{aligned}
u_m=(&{a_x}_m,{a_y}_m,{a_z}_m,\\
&{u_p}_1,{u_p}_2,{u_p}_3,{u_q}_1,{u_q}_2,{u_q}_3,{u_r}_1,{u_r}_2,{u_r}_3,\\
&{u_u}_1,{u_u}_2,{u_u}_3,{u_v}_1,{u_v}_2,{u_v}_3,{u_w}_1,{u_w}_2,{u_w}_3).
\end{aligned}
\end{equation*}
The main components are the measured accelerations \(a_m\) in the three respective axis. All the other components of the input vector are artificial inputs that are constantly set to 0. The reason for this is that for the rotational rates \(p,q,r\) and the wind components \(u_W,v_W,w_W\) no physical formulas in a manner required for equation \ref{StateEquation} can be derived. The proposed procedure is based on the Estimation-Before-Modeling method that was introduced in \cite{SriJayantha.1988} and the details can again be found in \cite{JoachimSiegel.25.04.2017} on pages 23-24. Due to the relevance of the variables \(p,q,r,u_W,v_W\) reconstructed versions of them are very beneficial. Therefore, these variables are added to the output vector \(y\). To be able to develop an output equation \(g\) for these variables, \(p,q,r,u_W,v_W,w_W\) also have to be added to the system states and consequently a state equations have to be developed for them. However, no direct physical relation exist that fulfills the requirement of the state equation, see \cite{JoachimSiegel.25.04.2017} page 23. The idea of the state equation based on the Estimation-Before-Modeling method for \(p\) is given by

\begin{equation}\label{GaussMarkovEquationp}
\left(\begin{array}{c}
\dot{p}\\\ddot{p}\\\dddot{p}
\end{array}\right)
=
\left(\begin{array}{ccc}
0&1&0\\0&0&1\\0&0&0
\end{array}\right)
\cdot
\left(\begin{array}{c}
	p\\\dot{p}\\\ddot{p}
\end{array}\right)
+
\left(\begin{array}{c}
{u_p}_1-{w_p}_1\\{u_p}_2-{w_p}_2\\{u_p}_3-{w_p}_3
\end{array}\right).
\end{equation}

From Equation \ref{GaussMarkovEquationp} it can be seen why the artificial constant inputs \({u_p}_1=0,{u_p}_2=0,{u_p}_3=0\) are required. The variable \(w_W\) was chosen to be not part of the output vector, however as a state influences several other output vector components.

\begin{figure}[ht]
	\centering
	\includegraphics[width=0.8\textwidth]{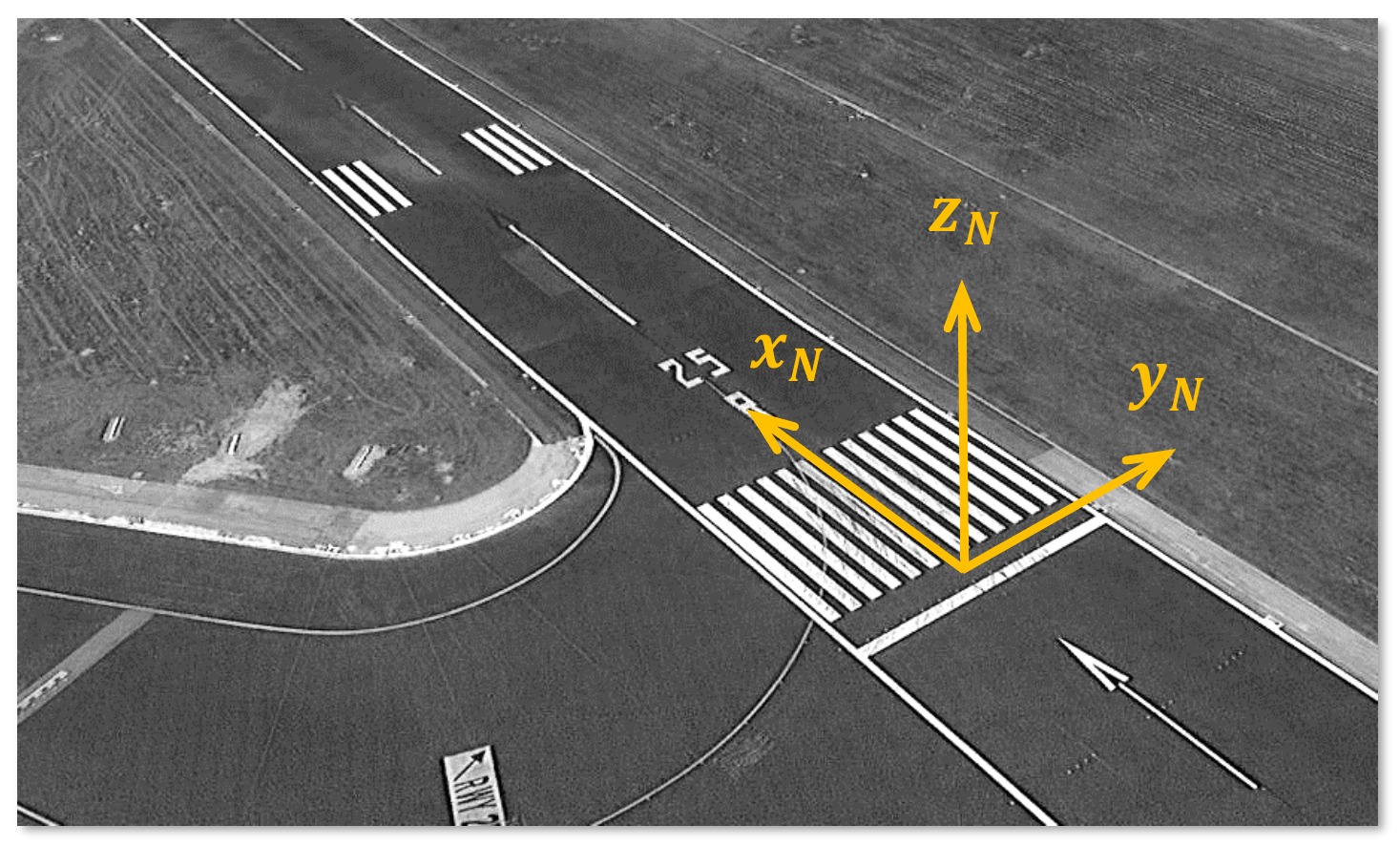}
	\caption{Local navigation frame, source: Figure 3-2 of \cite{Siegel.18.11.2015}}
    \label{NFrame}
\end{figure}

The state vector \(x\) comprises
\begin{equation}\label{RTSStateVector}
\begin{aligned}
x=(&(u_K)_B,(v_K)_B,(w_K)_B,\\
&\phi,\theta,\psi,\\
&x_N,y_N,z_N,\\
&p,\dot{p},\ddot{p},q,\dot{q},\ddot{q},r,\dot{r},\ddot{r},\\
&(u_W)_O,(\dot{u}_W)_O,(\ddot{u}_W)_O,\\
&(v_W)_O,(\dot{v}_W)_O,(\ddot{v}_W)_O,\\
&(w_W)_O,(\dot{w}_W)_O,(\ddot{w}_W)_O).
\end{aligned}
\end{equation}
The first three components of \(x\) describe the kinematic speed of the aircraft given in the three axes of the body fixed \(B\) frame. The attitude is described by the Euler angles \(\phi,\theta,\psi\) that are used to transfer the North-East-Down (NED) frame \(O\) into the body fixed \(B\) frame, see \cite{Zipfel.2014}. \(x_N,y_N,z_N\) describe the position of the aircraft in a local navigation frame \(N\), see Fig.\ \ref{NFrame}. Its origin is defined to be the intersection of the runway threshold and the runway center line. Its \(x\) axis points along the runway centerline and its \(z\) axis vertically upwards. The \(y\) axis points to the right and  complements an orthogonal frame. \(p,q,r\) are the rotational rates of the aircraft and their first and second order derivatives are added again based on the Estimation-Before-Modeling method of \cite{SriJayantha.1988}. The same holds for the three components of the wind speed given in the NED frame \(O\).

The output vectors \(y\) and \(y_m\) are given by
\begin{equation}\label{RTSOutputVector}
\begin{aligned}
y_m=(&{V_{GND}}_m,\dot{h}_m,{\chi_K}_m,\\
&\phi_m,\theta_m,\psi_m,\\
&{{x_N}}_m,{{y_N}}_m,\\
&{h_{BARO}}_m,{h_{RALT}}_m,\\
&{\delta_{LLZ,DDM}}_m,{\delta_{GS,DDM}}_m,\\
&p_m,q_m,r_m,\\
&{V_A}_m,{\alpha_A}_m,\\
&{(u_W)_O}_m,{(v_W)_O}_m).
\end{aligned}
\end{equation}
They consist of the ground speed \(V_{GND}\), the vertical speed \(\dot{h}\), and kinematic track angle \(\chi_K\). Furthermore, the Euler angles \(\phi,\theta,\psi\) describe the attitude and the positions \(x_N,y_N\) based on recordings of the Global Positioning System (GPS). The altitude is given twofold, the barometric altitude, and the height above ground recorded by the radio altimeter. Within this paper it is assumed that the aircraft was flying an Instrument Landing System (ILS) approach and the related signals are available in the FDM data. The deviations of the aircraft with respect to the ILS are given by \(\delta_{LLZ,DDM}\) in the horizontal plane and \(\delta_{GS,DDM}\) in the vertical plane. Their unit is Difference in the Depth of Modulation (DDM). In addition to the rotational rates \(p,q,r\) and the horizontal wind components \((u_W)_O\) and \((v_W)_O\), also the aerodynamic speed \(V_A\) and aerodynamic angle of attack \(\alpha_A\) are given.

The bias and scale factor vector \(\Theta\) comprises
\begin{equation}\label{ThetaVector}
\begin{aligned}
\Theta=(&b_x,b_y,b_z,\\
&b_p,b_q,b_r,\\
&b_{h_{BARO}},s_{h_{BARO}},\\
&b_{\chi}).
\end{aligned}
\end{equation}

The biases for the accelerations \(a_x,a_y,a_z\) are denoted by \(b_x,b_y,b_z\) and are subtracted from the measured values, e.g.\ \(a_x = {a_x}_m-b_x\). The biases for the rotational rates \(p,q,r\) are denoted by \(b_p,b_q,b_r\) and again subtracted from the measured values, e.g.\ \(p = p_m-b_p\). The barometric altitude is modeled with both, bias and scale factor corrections \ \({h_{BARO}}_m = s_{h_{BARO}}\cdot h_{BARO} + b_{h_{BARO}}\). Finally, there is a bias \(b_\chi\) for the kinematic track angle \(\chi_K\) which is again subtracted from the recorded value.

The model contains too many equations to give any detail within this paper and the reader is again referred to \cite{JoachimSiegel.25.04.2017}. The two output vector components regarding ILS deviations \(\delta_{LLZ,DDM}\) and \(\delta_{GS,DDM}\) are exceptional for this implementation and bring a lot of benefit due to a high data accuracy of these signals compared to other FDM variables. As one example of an output equation that is part of the model, the one for \(\delta_{LLZ,DDM}\) is described here.

\begin{figure}[ht]
	\centering
	\includegraphics[width=0.9\textwidth]{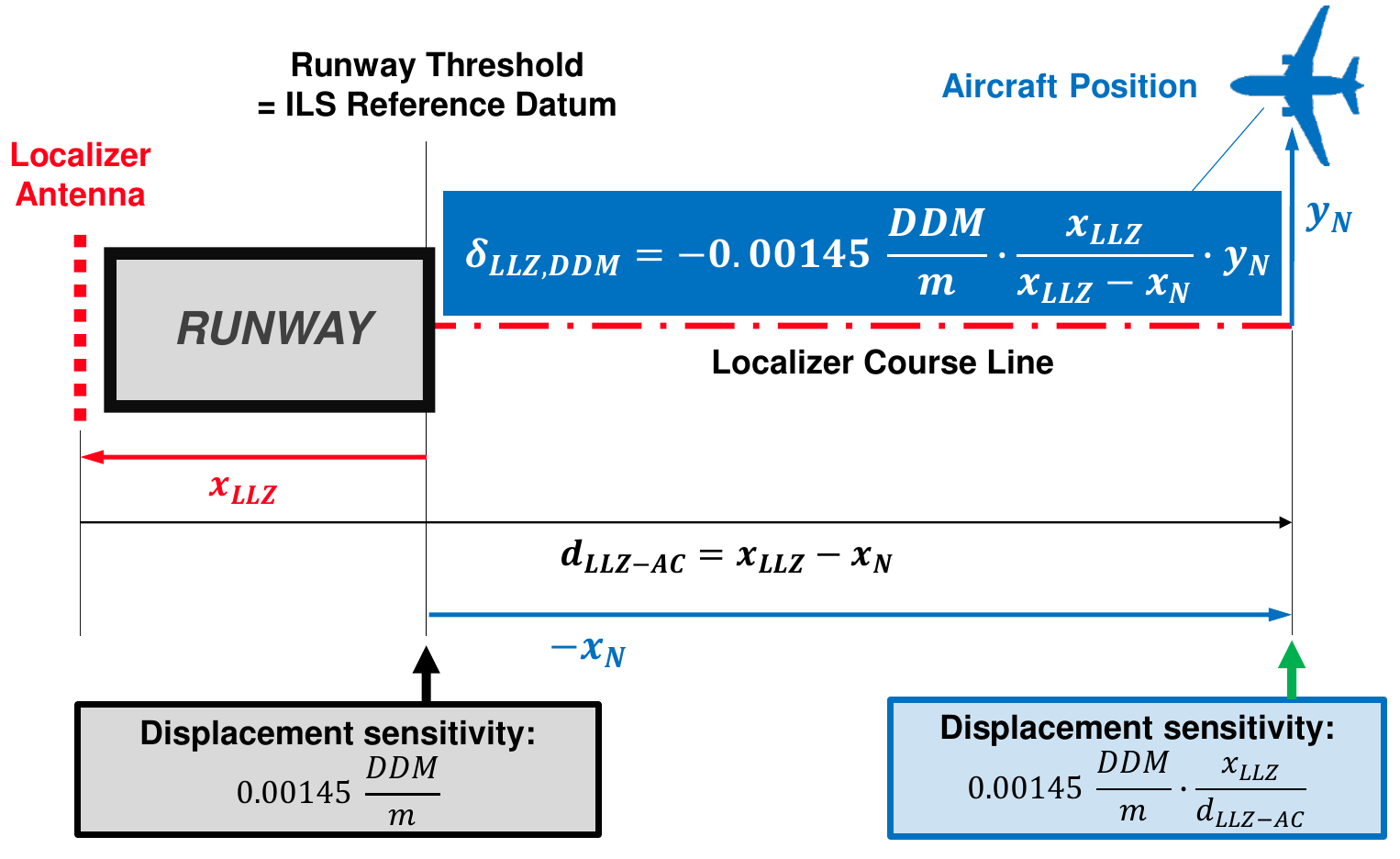}
	\caption{Output equation for the localizer deviation signal, source: Figure 2-14 of \cite{JoachimSiegel.25.04.2017}}
    \label{LLZEquationFigure}
\end{figure}

The goal of the output equation \(g\) for the output component \(\delta_{LLZ,DDM}\) is to link the system states with \(\delta_{LLZ,DDM}\), see Equation \ref{OutputEquation}. For this equation, the system inputs \(u\) and the constant model parameters \(\Theta\) are not relevant. The relevant system states \(x\) are the aircraft position given by \(x_N\) and \(y_N\). According to \cite{ICAOAnnex10.2006}, the nominal displacement at the ILS reference datum (the runway threshold) shall be adjusted to 0.00145 DDM/m. The output equation can be derived using basic geometric considerations, see Fig.\ \ref{LLZEquationFigure}. It is given by
\begin{equation*}
\delta_{LLZ,DDM}=g(x_N,y_N)=-0.00145\cdot\frac{x_{LLZ}}{x_{LLZ}-x_N}\cdot y_N.
\end{equation*}
The parameter \(x_{LLZ}\) is a characteristic of the specific runway and the ILS and describes the longitudinal distance between the runway threshold and the localizer antenna, see Fig. \ref{LLZEquationFigure}.

In addition to the application of the RTS smoother framework that is discussed in chapter \ref{SmootherChapter}, a shifting of the trajectory of the landing aircraft based on the taxiway coordinates and specific assumptions that are well fulfilled in practice is conducted. Indicating the details is again out of the scope of this paper and the reader is referred to \cite{JoachimSiegel.25.04.2017}.

\section{Kalman Filter, Extended Kalman Filter, and the RTS Smoother}\label{SmootherChapter}
The goal of this chapter is to give an introduction to the RTS smoother, which is an extension of the Extended Kalman Filter (EKF). The historic origin of this theory is given by Rudolf Emil Kálmán in \cite{Kalman.1960} for the linear case and has been extended to the non-linear case in \cite{Schmidt.1976}. The characteristics of offline data analysis, as it is the case in FDM, allows to take data of the past and the future of a certain time step \(t\) into account. These additional possibilities lead to the concept of the Rauch-Tung-Striebel (RTS) smoother that was introduced in \cite{Rauch.1965}. A thorough introduction to the underlying theory that is often referred to within this paper is given by \cite{Jategaonkar.2015b}. The RTS smoother and the associated physical model (see chapter \ref{PhysicalModelChapter}) that are implemented at the Institute of Flight System Dynamics and have been primarily developed in \cite{JoachimSiegel.25.04.2017} and \cite{Siegel.18.11.2015}.

\begin{figure}[ht]
	\centering
	\includegraphics[width=0.8\textwidth]{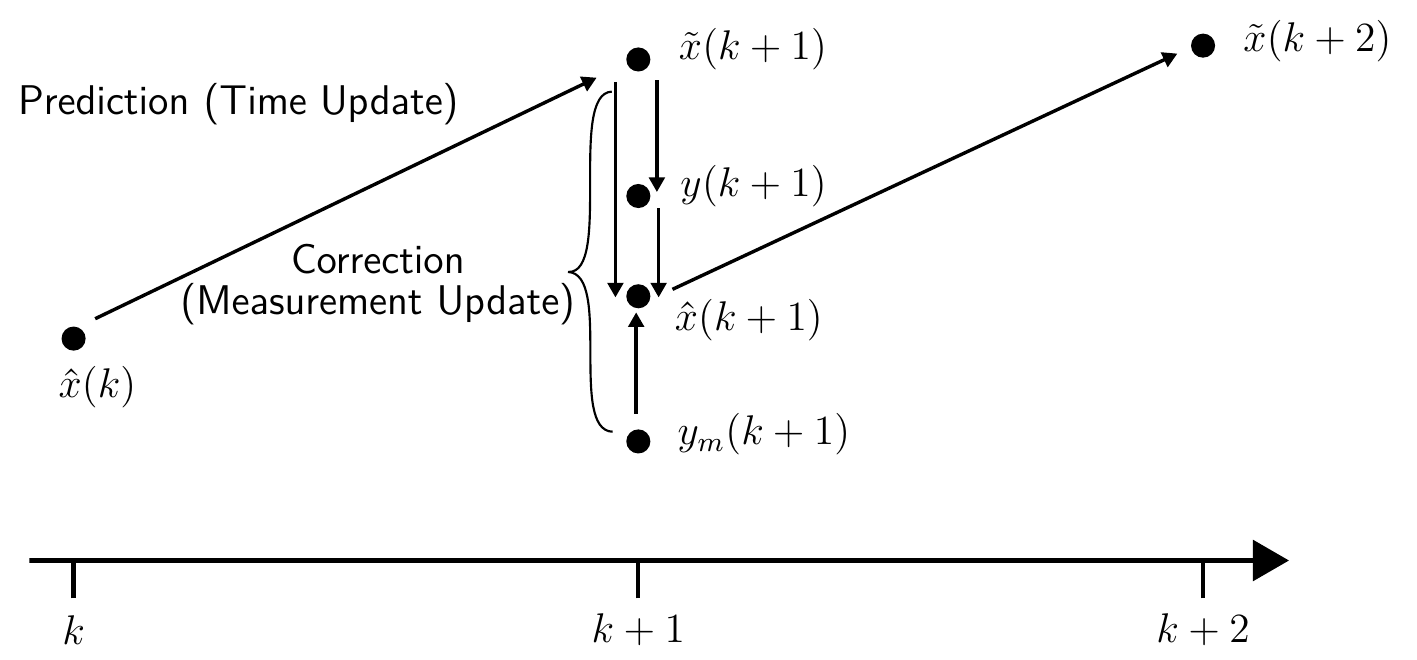}
	\caption{Kalman filter principle based on Figure F1 of \cite{Jategaonkar.2015b}, page 598}
    \label{KalmanFilterPrinciple}
\end{figure}

The Kalman filter and its generalized version for the non-linear case EKF consist of prediction steps and correction steps, see Fig.\ \ref{KalmanFilterPrinciple}. Values related to the prediction step are denoted by the superscript \myTilde\ and the corrected states denoted by \^{}. Based on the corrected state \(\hat{x}(k)\) at time step \(k\), the state \(\tilde{x}(k+1)\) at time step \(k+1\) is predicted based on the physical model. Subsequently, the measured output \(y_m(k+1)\) at time step \(k+1\) is used to correct \(\tilde{x}(k+1)\) which leads to \(\hat{x}(k+1)\). The true output \(y(k+1)\) at time step \(k+1\) is unknown. This process is repeated iteratively until the end of the considered period.

This iterative process of the EKF is also referred to as the forward pass of the RTS smoother. Once the end of the time interval is reached, the RTS smoother adds a further backward pass in which the time interval is passed through from the end to the begin of the interval taking the reversed system dynamics into account. This leads to a three step approach, a forward pass, a backward pass, and a final pass to weigh and combine the results from the two passes. However, \cite{Rauch.1965} proposes to combine the second and third pass.

Data is recorded with a certain frequency and not continuously. The notation \(v_k = v(t_k)\) is often used in the following for \(v\) and other variables. Both noise terms, the input measurement noise \(w\) and the output measurement noise \(v\), are modeled as zero mean Gaussian white noise. In particular, the noise terms are supposed to satisfy the following relations
\begin{gather}
\mathbb{E}(w_k)=0\\
\mathbb{E}(v_k)=0 \label{eq:Ev}\\
\mathbb{E}(w_k\cdot w_l^T)=
\begin{cases}
Q_k&\text{for } k=l\\
0&\text{for } k\neq l
\end{cases}\\
\mathbb{E}(v_k\cdot v_l^T)=
\begin{cases}\label{RDefinition}
R_k&\text{for } k=l\\
0&\text{for } k\neq l
\end{cases} \\
\mathbb{E}(w_k\cdot v_l^T)=0\text{ for all } k,l\in\mathbb{N}. 
\end{gather}

The measurement noise covariance matrices \(R_k\) is central for this paper. \(R_k\) is directly occurring in the correction step
\begin{gather}
	K_{k+1}=\tilde{P}_{k+1}\cdot C_{k+1}^T\cdot (C_{k+1}\cdot\tilde{P}_{k+1}\cdot C_{k+1}^T+R_{k+1})^{-1}\label{KalmanGainWithR}\\
	\begin{aligned}\label{StateCovarianceWithR}
	\hat{P}_{k+1}&=(I-K_{k+1}\cdot C_{k+1})\cdot\tilde{P}_{k+1}\\=(I-K_{k+1}\cdot C_{k+1})\cdot&\tilde{P}_{k+1}\cdot(I-K_{k+1}\cdot C_{k+1})^T+K_{k+1}\cdot R_{k+1}\cdot K_{k+1}^T.
	\end{aligned}
\end{gather}

The sequence of matrices \(K_k\) are the Kalman gain that weighs the predictions and the measurements. The matrices \(K_k\) are derived such that the associated state covariance matrices \(P_k\) are minimized. \(I\) denotes the identity matrix in the corresponding dimension. The matrix \(C_k\) is the linearization of the output equation

\begin{equation*}
	\left.C_{k+1}=\frac{\partial g(x,u,\Theta)}{\partial x}\right\vert_{x=\tilde{x}_{k+1},u={u_m}_k}.
\end{equation*}

The reader is referred to the literature for all the formulas of the RTS smoother framework and their details. Only the main concepts and main equations necessary to understand the ideas and algorithms of this paper are given explicitly. The derivation of the Kalman filter and EKF can for example be found in Appendix F, pages 597-607 of \cite{Jategaonkar.2015b}. The derivation of the RTS smoother can be found in \cite{Rauch.1965}.

\section{Smoothing Quality Measure}\label{QualityMeasureChapter}
FDM analyses have to be automatized as much as possible. Due to the vast number of flights of a major airline it is simply not feasible to manually investigate each individual flight. Consequently, this principle is also relevant for the reconstruction of landings using the RTS smoother.

In \cite{JoachimSiegel.25.04.2017}, a high number of flights was analyzed and a Smoothing Quality Measure (SQM) was introduced. The goal was to automatically detect flights where the RTS smoothing could not generate reasonable results, i.e.\ the smoothing quality is low. This might happen for instance due to severely corrupted flight data.

The SQM that was developed in Chapter 2.1.9 of \cite{JoachimSiegel.25.04.2017} has been modified within this project to take the characteristics of the time varying measurement noise covariance matrices into account. For each time step \(k=1,\dotsc,N\) with \(N\) being the number of total time steps of the considered time interval,
\begin{equation*}
\varepsilon_{k}=y_{m,k}-\tilde{y}_k\in\mathbb{R}^n
\end{equation*}
is the difference between the measured and the predicted output. The theoretical values of the residual covariance matrix \(S_k\) for each time step \(k\) can be computed as
\begin{equation*}
S_{k}=Cov(\varepsilon_k)=C_{k}\cdot\tilde{P}_{k}\cdot C_{k}^T+R_{k}.
\end{equation*}

Furthermore, for any component \(i=1,\dotsc,n\) of the output vector 
\begin{equation*}
r_i=\frac{1}{N}\cdot \sum_{k=1}^N\frac{(\varepsilon_{k,i}-\bar{\varepsilon}_i)^2}{S_{k,ii}}
\end{equation*}
is defined. Thereby, the numerator corresponds to the empirical variance of the prediction error \(\varepsilon\) considering all time steps \(N\). The denominator describes its theoretical value based on the RTS smoother theory. In the optimal case, the empirical variance and the expected variance coincide and so resulting in \(r_i\) around 1. Abnormal situations may lead to \(r_i\) severely deviating from 1.

Finally, the values \(r_i\) for all \(n\) output vector components are aggregated to the SQM using the geometrical mean
\begin{equation*}
\text{SQM} = \left[\prod_{i=1}^n r_i\right]^{\frac{1}{n}}.
\end{equation*}

The interpretation for \(r_i\) also holds for SQM. Flights with a regular smoothing will generate a SQM close to 1. If the prediction errors outweigh the expected prediction errors, the SQM gets bigger. An SQM greater 10 can be considered as abnormal.

In this article the SQM is used for verification whether the integration of time varying measurement noise covariance matrices is beneficial or not.

\section{Measurement Noise Covariance Estimation}\label{NoiseCharacterization}

The idea is straight forward and also mentioned for time varying filters on page 181 of \cite{Jategaonkar.2015b}. Taking Equations \ref{OutputEquationWithErrors} and  \ref{RDefinition} together with the noise $v_k = y_m(k)-g(x(k),u_m(k),\Theta)$ gives
\begin{equation*}
R_k = \mathbb{E}\bigl(v_k\cdot v_k^\top\bigr),
\end{equation*}
which is an $n \times n$ matrix where the $(i,j)$ entry corresponds to the covariance between the $i$th and $j$th component of $v_k$.
For the unknown values \(g(x(k),u_m(k),\Theta)\) the outputs reconstructed in an initial smoother iteration 
\begin{equation*}
\hat{y}_k=g(\hat{x}(k),u_m(k),\hat{\Theta})
\end{equation*}
with constant covariance matrix $R_k=R$ are used. Furthermore, the estimated residuals vector is given by
\begin{equation}\label{EstimatedResiduals}
\hat{v}_k=y_m(k)-\hat{y}_k.
\end{equation}

To estimate $R_k$, the output covariance matrix at time step $k$, we use a moving average approach as proposed in \citet{Yin.2010}. More specifically, the estimate $\widehat R_k$ is defined as
\begin{align} \label{eq:Rhat}
\widehat R_k = \sum_{t = 1}^N [v_t-\widehat m_t]\cdot[v_t-\widehat m_t]^\top\cdot w_b(t, k),
\end{align}
where
\begin{align*}
\widehat m_k =  \sum_{t = 1}^N v_t\cdot w_b(t ,k), \qquad w_b(t, k) = \frac{e^{-\frac{(t - k)^2}{2 b}}}{\sum_{t = 1}^N e^{-\frac{(t - s)^2}{2 b}}}.
\end{align*}
The function $w_b$ assigns weights to all time steps such that time points closer to $k$ have a larger influence and time points further away from $k$ have only a small influence. The value $b$ controls the degree of smoothing and is set to $50$. Note that $\widehat m_k$ is an estimate of $\mathbb{E}(v_k)$. Although this expectation is assumed to be zero in RTS smoother, see Equation \eqref{eq:Ev}, we often observe significant deviations from zero in practice. Thus, $\widehat m(t)$ is used as a correction term in Equation \eqref{eq:Rhat}. 

\begin{figure}[ht]
	\centering
	\includegraphics[width=0.8\textwidth]{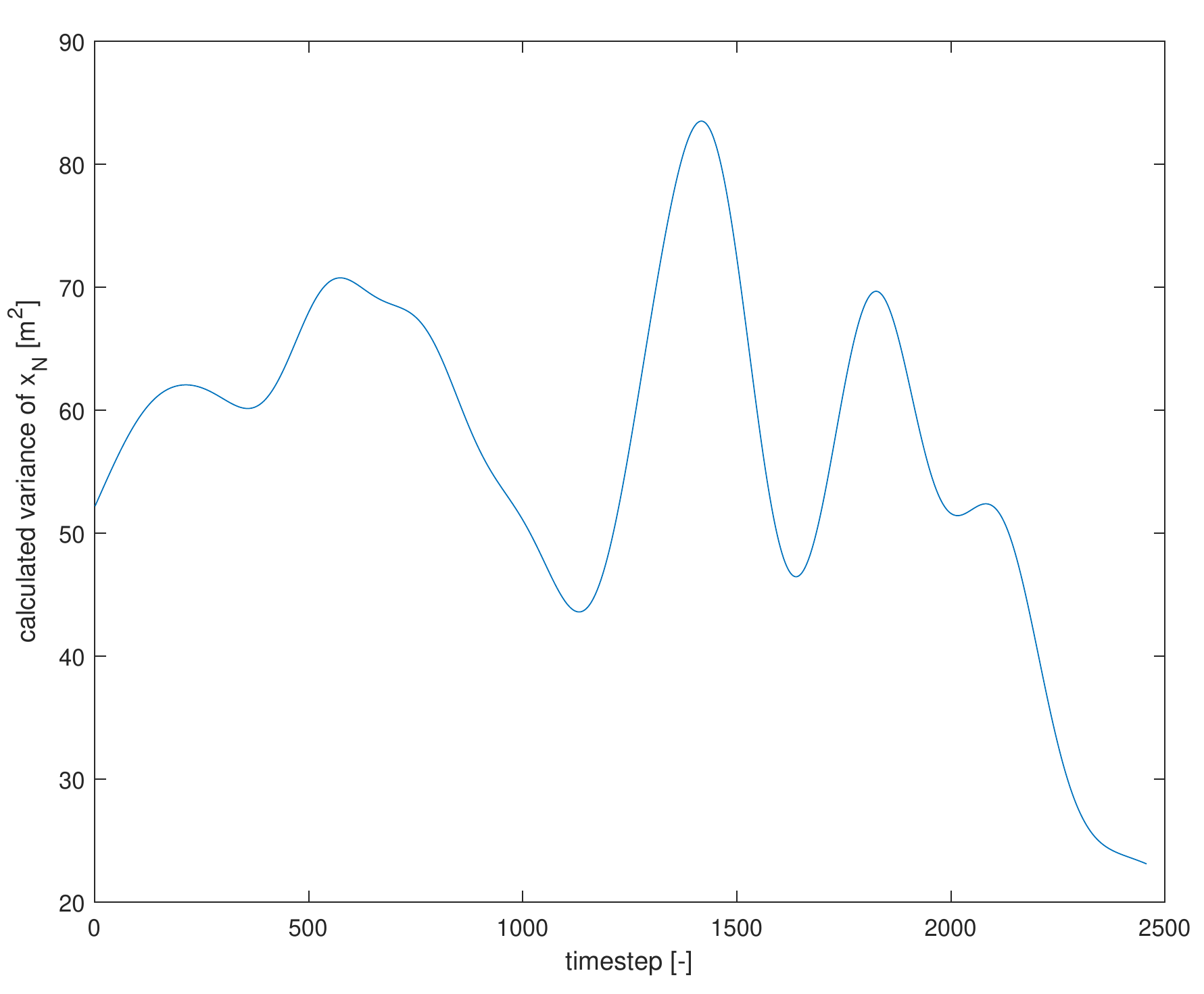}
	\caption{Time varying measurement noise variance for \(x_N\)}
    \label{Covariance}
\end{figure}

An exemplary estimate of the time varying variance of $x_N$ is shown in Fig. \ref{Covariance}, where one clearly observes periods of higher and smaller variability. 

% The off-diagonal entries of $\widehat R(t)$ correspond to estimated covariances. In general, they are non-zero, which may reflect dependence between measurement errors. But often non-zero entries are merely a consequence of estimation errors. In preliminary experiments we found that small off-diagonal entries destabilize RTS smoother. We account for this by a thresholding approach: whenever the absolute value of the correlation between a pair of parameters is less than a prespecified threshold, the corresponding entry in the covariance matrix is set to zero. We then run the smoother with different correlation limits and pick the one that improves the SMQ the most.

\section{Integration of Time Varying Covariance Matrices in the RTS Smoother}\label{IntegrationNoiseRTSChapter}
For an initial smoother iteration, a constant measurement noise covariance matrix \(R=R_k\) is chosen based on expert judgment. This matrix \(R\) is assumed to be a diagonal matrix and the associated values can be found in \cite{JoachimSiegel.25.04.2017}. The reconstructed outputs of this smoother iteration can be used to obtain an estimation of the true measurement noise characteristic that is more accurate than the defined matrix. In addition, potential time varying noise characteristics can be taken into account using time varying noise covariance matrices. The details were given in chapter \ref{NoiseCharacterization}.

It turns out that taking the sequence of full matrices \(\widehat R_k\) of chapter \ref{NoiseCharacterization} is not stable enough. Therefore a limit for the off-diagonal entries based on the related correlation has been introduced. The diagonal entries are directly taken as calculated in chapter \ref{NoiseCharacterization}. % One exception is the variable \(h_{WGS84}\). For the flights considered in this paper there are no recordings of this value available. The values of the associated column and row of \(R_k\) are set to 0, except for the diagonal element for \(h_{WGS84}\) in \(R_k\) for which the value of the first smoother iteration is used. Otherwise the matrix gets singular causing numerical problems.

\begin{figure}[H]
	\centering
	\includegraphics[width=0.9\textwidth]{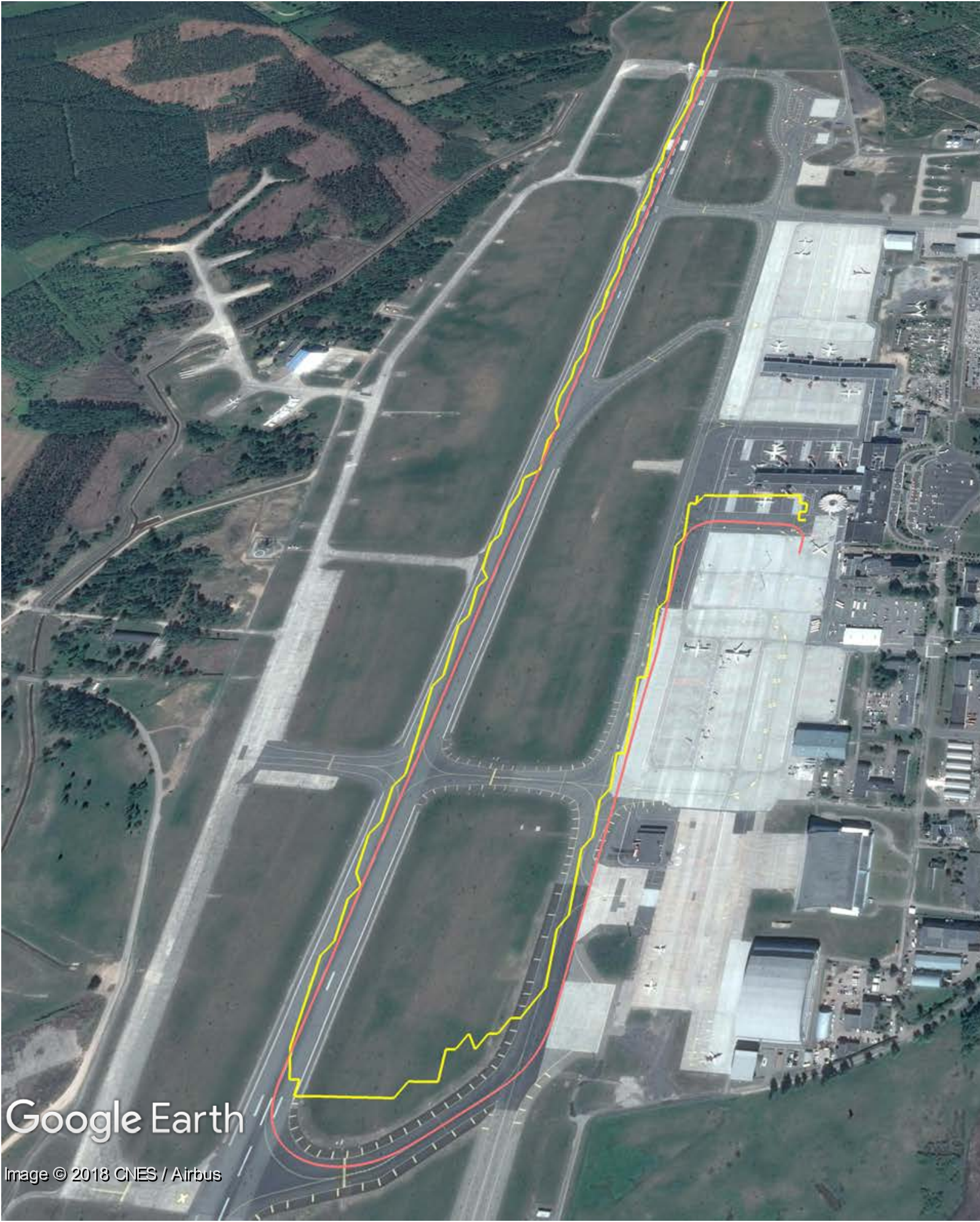}
	\caption{Reconstructed position - correlation limit of the measurement noise covariance matrices \(R_k\) is 0.1}
    \label{GooglePlot}
\end{figure}

Reconstructed time series for all states \(x\) and output variables \(y\) are retrieved. Within this paper the reconstructed position is used for visualization. In Fig.\ \ref{GooglePlot}, the yellow curve shows the raw position data and the red curve shows the reconstructed trajectory after the second smoother iteration with a limit of the \(R_k\) off-diagonal entries based on a correlation of 0.1.

To investigate the influence of the second smoother iteration with incorporated time varying measurement noise covariance matrices \(R_k\), the SQM described in chapter \ref{QualityMeasureChapter} is used. Within this paper, 24 flights are considered and the results of the SQM values are summarized in Table \ref{SQMTable}.

\definecolor{darkspringgreen}{rgb}{0.09, 0.45, 0.27}

\begin{table}
\centering
\begin{tabular}{ c|c|c|c|c|c } 
Flight ID & 1\textsuperscript{st} Iter. & 2\textsuperscript{nd} Iter., Limit 0.1 & 2\textsuperscript{nd} Iter., Limit 0.4 & 2\textsuperscript{nd} Iter., Limit 0.6 & 2\textsuperscript{nd} Iter., Limit 0.8 \\ 
\hline
1 & 0.24 & 0.45\cellcolor{gray} & 20.3 & 0.52 & 0.53\cellcolor{darkspringgreen}\\ 
\hline
2 & 0.18 & - & \cellcolor{darkspringgreen}0.49 & 0.44 & 0.45\\ 
\hline
3 & 0.36 & - & 15.4 & 0.61 & \cellcolor{darkspringgreen}0.62\\ 
\hline
4 & 0.46 & 0.58\cellcolor{gray} & \cellcolor{darkspringgreen}0.81 & 0.76 & 0.76\\ 
\hline
5 & 0.44 & \cellcolor{darkspringgreen}0.86 & - & 0.65 & 0.65\\ 
\hline
6 & \cellcolor{darkspringgreen}0.39 & 0.90\cellcolor{gray} & - & 109 & \(5\cdot10^{7}\)\\ 
\hline
7 & \cellcolor{darkspringgreen}0.79 & - & - & 0.65 & 0.65\\ 
\hline
8 & 0.41 & - & 0.82\cellcolor{gray} & 0.64 & \cellcolor{darkspringgreen}0.64\\ 
\hline
9 & 0.42 & 0.59\cellcolor{gray} & 0.44\cellcolor{gray} & \cellcolor{darkspringgreen}0.74 & 0.74\\ 
\hline
10 & 0.51 &- & - & 9.1 & 0.69\cellcolor{darkspringgreen}\\ 
\hline
11 & 0.51 & - & - & - &\cellcolor{darkspringgreen} 0.70\\ 
\hline
12 & 0.35 & - & 1900\cellcolor{gray} & \cellcolor{darkspringgreen}0.71 & 0.61\\ 
\hline
13 & 0.21 & 0.43 & - & 0.51 & 0.51\cellcolor{darkspringgreen}\\ 
\hline
14 & 0.79\cellcolor{darkspringgreen} & 0.60\cellcolor{gray} & 1.83 & 1.64 & 1.64\\ 
\hline
15 & 0.31 & 11.1 & \cellcolor{darkspringgreen}0.59 & 0.53 & 0.53\\ 
\hline
16 & 0.67 & - & \cellcolor{darkspringgreen}0.80 & 0.65 & 0.59\\ 
\hline
17 & 0.20 & 0.61\cellcolor{gray} & - & 0.48 & \cellcolor{darkspringgreen}0.48\\ 
\hline
18 & 0.37 & - & \cellcolor{darkspringgreen}0.67 & 0.65 & 0.65\\ 
\hline
19 &\cellcolor{darkspringgreen} 0.93 & 0.50\cellcolor{gray} & - & 1.85 & 1.85\\ 
\hline
20 & 0.23 & 0.47\cellcolor{gray} &\cellcolor{darkspringgreen} 0.53 & 0.51 & 0.49\\ 
\hline
21 & 0.24 & \cellcolor{darkspringgreen}0.67 & 0.63 & 0.55 & 0.55\\ 
\hline
22 & 0.37 & 0.98\cellcolor{gray} & - &\cellcolor{darkspringgreen} 0.62 & 0.62\\ 
\hline
23 & 0.31 & 60.0 & 0.88\cellcolor{gray} & \cellcolor{darkspringgreen}0.61 & 0.61\\ 
\hline
24 & 0.38 & 0.62\cellcolor{gray} & \cellcolor{darkspringgreen}0.76 & 0.67 & 0.67\\ 
\end{tabular}
\caption{SQM values for different iterations and correlation limits of the measurement noise covariances}
\label{SQMTable}
\end{table}

As described in chapter \ref{QualityMeasureChapter}, the value of the SQM considered as optimal within this paper is 1. Within Table \ref{SQMTable}, the value closest to 1 has been highlighted for any flight in green color. In addition, SQM values for which not the entire considered parameters were constructed are highlighted in gray color. The implemented smoother can handle these situations and therefore valid SQM values are returned. However for the purpose of this paper, it was chosen that always the entire parameters are considered.

For the considered 24 flights, the optimal SQM value can be found in a second iteration with a specific covariance limit in 20 cases (see cells highlighted in green). This corresponds to a percentage of 83\% which shows the justification of this method. However, values far away from 1 or missing values show that the second iteration and the related computations are also prone to further errors. The best reconstruction strategy therefore might be, to conduct second iterations with several correlation limits and subsequently choose the version among the first iteration and all the second iterations with the SQM value closest to 1.

\section{Validity of Assumptions in the (improved) RTS Smoother} \label{DependenceInvestigationChapter}

We now assess the validity of some of the assumptions on $v_k$ discussed in  chapter \ref{SmootherChapter}. On that account, we visually illustrate statistical properties of the estimated residual process $\widehat v_k$ and discuss possibilities for further improvement.

\subsection{Constant mean}

\begin{figure}[ht]
	\centering
	\includegraphics[width=\textwidth]{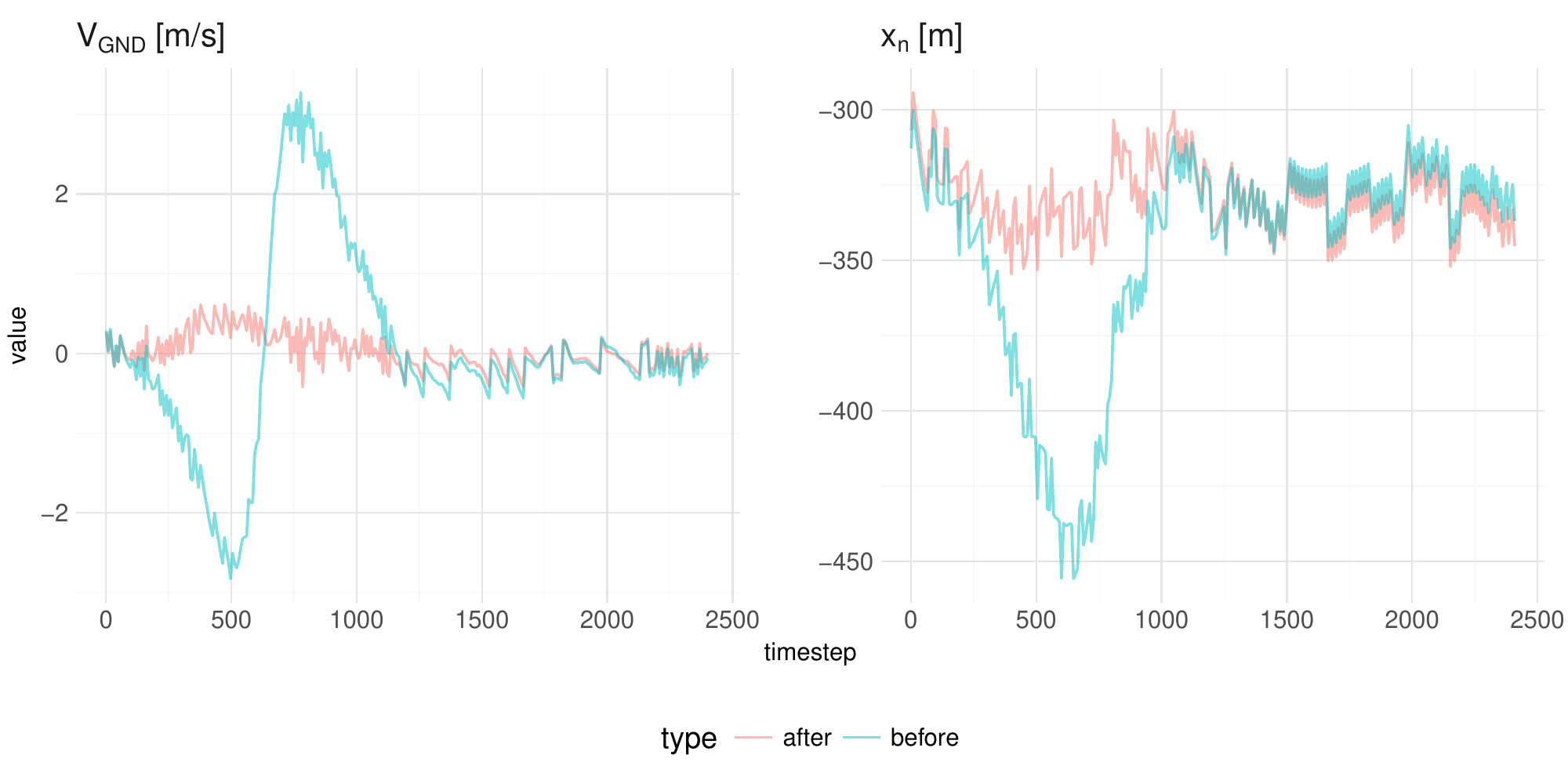}
	\caption{Time series of estimated residuals (see equation \ref{EstimatedResiduals}) for variables \(V_{GND}\) and \(x_N\)}
    \label{ImprovedAssumptions}
\end{figure}

Recall from equations \eqref{eq:Ev} and \eqref{RDefinition} that $v_k$ is assumed to be a white noise process. As an illustration, Fig.\ \ref{ImprovedAssumptions} shows the residuals for the variables \(V_{GND}\) and \(x_N\).
The estimated $\hat{v}_k$ process from the initial smoother run with constant covariance matrix (blue lines) strongly deviates from this assumption: the mean is far from constant and the series is highly autocorrelated. In contrast, the estimates process from the second smoother iteration with time varying covariance matrix (red lines) is visibly more truthful to these assumptions, although there is still room for improvement. But in contrast to dynamic covariances, these two assumptions are only indirectly influenced by the parametrization of the smoother.

\subsection{Gaussianity}

Another assumption made in chapter \ref{SmootherChapter} is that the distribution of $v_k$ is (multivariate) Gaussian. This assumption can be split into two parts: i) the components of $v_k$ are marginally Gaussian; ii) the dependence between components follows a Gaussian copula \citep[see, e.g.,][]{nelsen2007introduction}.

\subsubsection*{Marginal distributions}

\begin{figure}[ht]
	\centering
	\includegraphics[width=\textwidth]{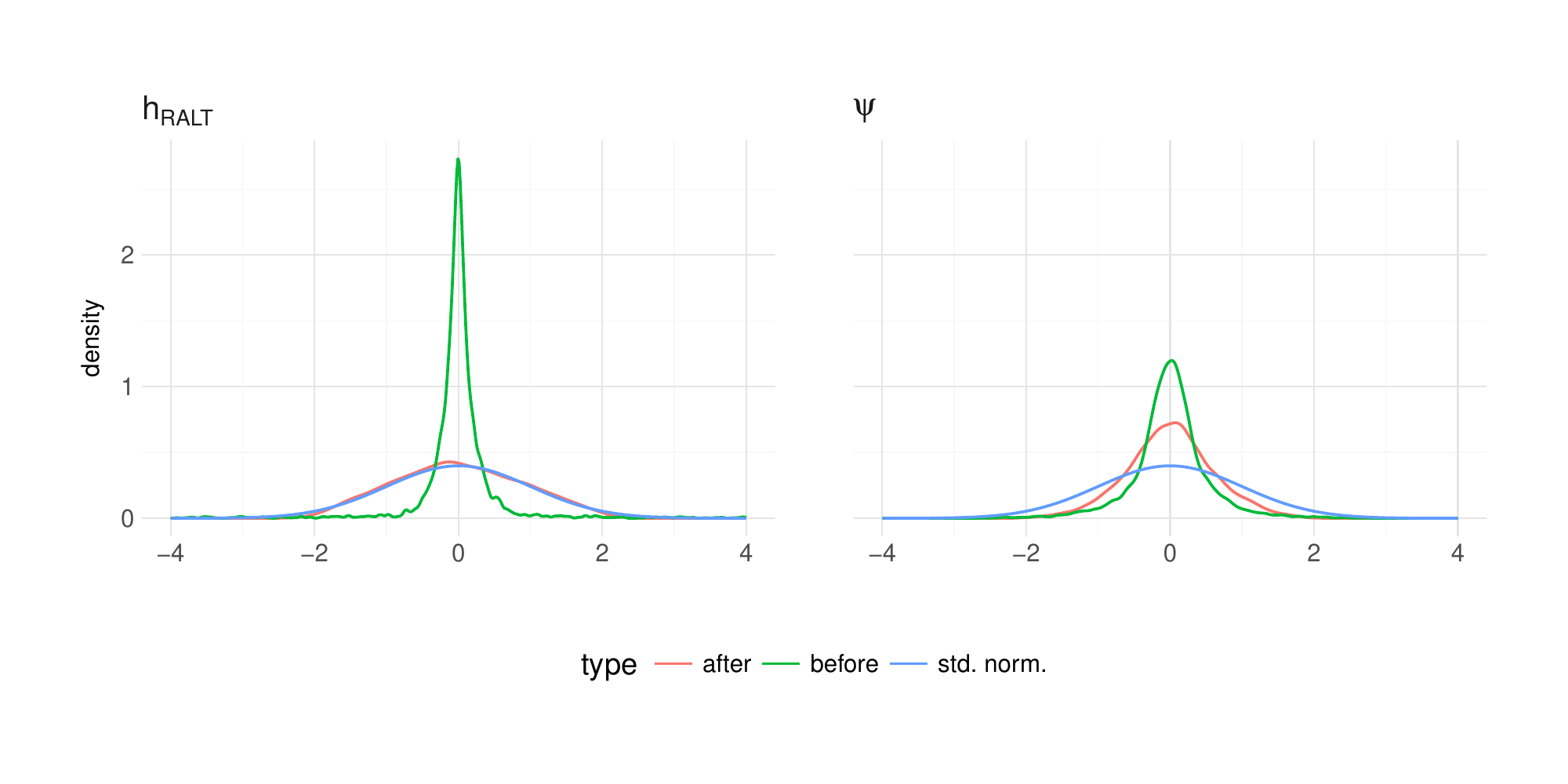}
	\caption{Kernel density plots for the standardized residual process $\widehat s_{k}$ (aggregated over all flights) which should follow a standard normal distribution.}
    \label{ImprovedAssumptions2}
\end{figure}

We shall first consider the marginal distributions. For each component $i$ of $\widehat v_k$ and every time step $k$, we construct standardized residuals $\widehat s_{k, i} = \widehat v_{k, i} / (\widehat R_{ii})^{1/2}$, where $\widehat R_{ii}$ is the estimated standard deviation at time $k$. If the assumptions are valid, $\widehat s_{k, i}$ should follow a standard normal distribution. 

Fig.~\ref{ImprovedAssumptions2} shows kernel density plots for two exemplary variables along with the density of a standard normal distribution (blue line). Results after the first smoother iteration are shown in green, results after the second iteration in red. For $h_{RALT}$ (left panel), the distribution for the first iteration is far off the standard normal, while the distribution for the second iteration almost perfectly matches it. For $\psi$ (right panel), the second iteration improves upon the first, but is still visibly off the assumed distribution. It has a sharper spike in the center and fatter tails, features that commonly arise in scale mixture of normal distributions (such as the Student $t$ distribution). It suggests that our time varying variance estimate may not have been adaptive enough and a smaller smoothing window could improve the results further.

\subsubsection*{Dependence analysis}

A useful diagnostic tool for the dependence between variables is the bivariate normalized contour plot \citep{nagler2017kdecopula}. In a first step, the variables are transformed to follow a standard normal distribution. If the dependence is Gaussian, the density contours of the transformed variables should resemble perfect ellipses. Any deviation from an ellipse indicates a deviation from normality. Most residual dependencies are rather weak and we will only focus on the strong relationships in what follows. Also, the second smoother iteration was not able to improve upon the first regarding the validity of Gaussian dependence, so only results from the second iteration will be shown.

% The main dependencies remaining in the residuals $y_m(k)-g(x(k),u_m(k),\Theta)$, see red lines of Fig.\ \ref{ImprovedAssumptions}, were investigated using copula models. The outputs reconstructed after the second smoother iteration were considered for the unknown values \(g(x(k),u_m(k),\Theta)\). Note that all dependencies should have been properly captured by \(R_k\) during the second smoother iteration so that ideally, no further dependencies in the residuals are present. All pairs with notable remaining dependence show a positive behavior, i.e.\ the higher the residual in the first variable, the higher it is also in the second variable. In case of an occurring negative dependence, a major revision of the physical model or the methods proposed in this paper to capture the dependencies and their integration into the RTS smoother framework would be necessary.

\begin{figure}[H]
	\centering
	\includegraphics[width=\textwidth]{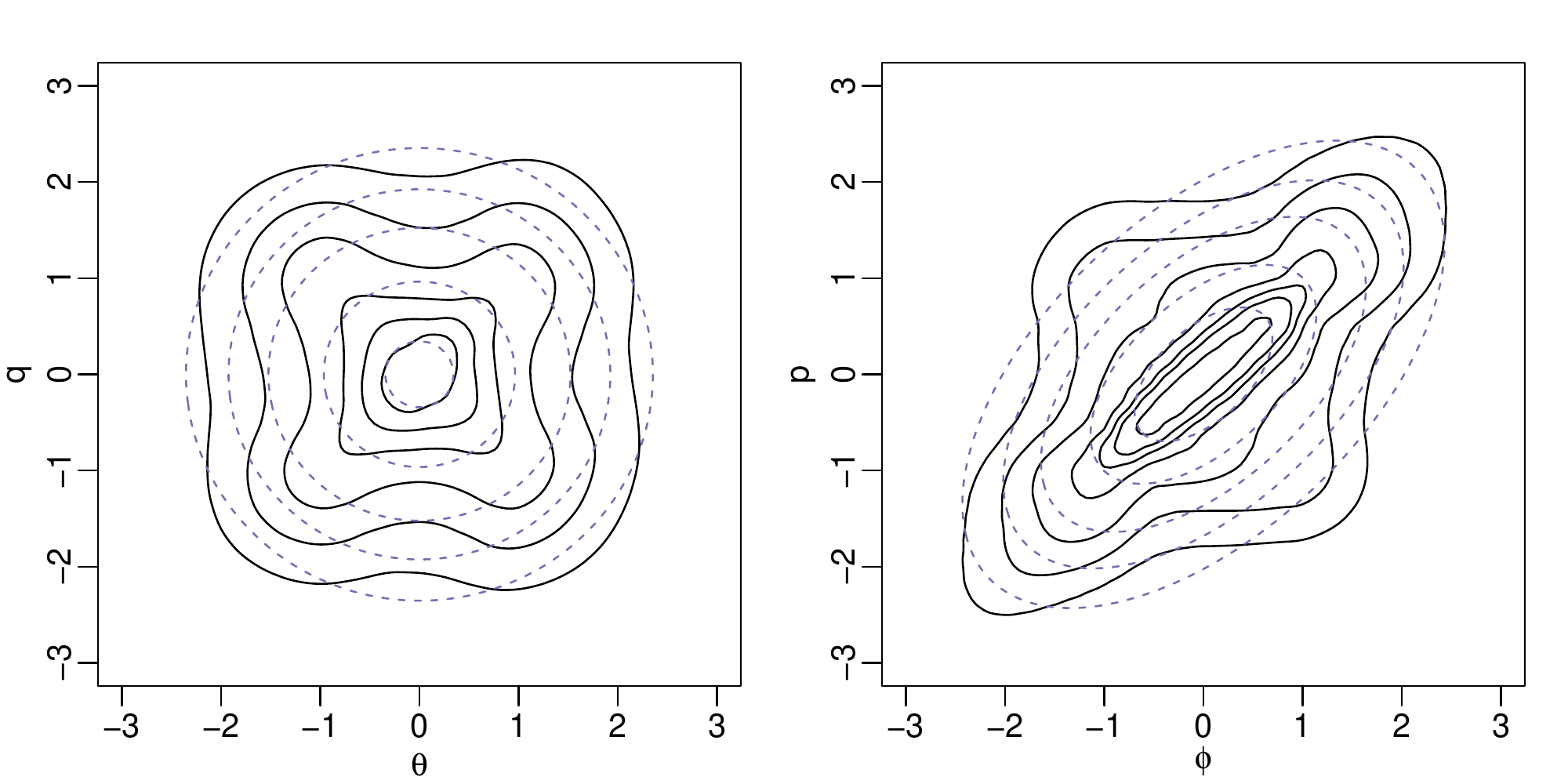}
	\caption{Copula dependence structures - attitude and rotational rates}
    \label{ContoursAttitude}
\end{figure}

\begin{figure}[H]
	\centering
	\includegraphics[width=\textwidth]{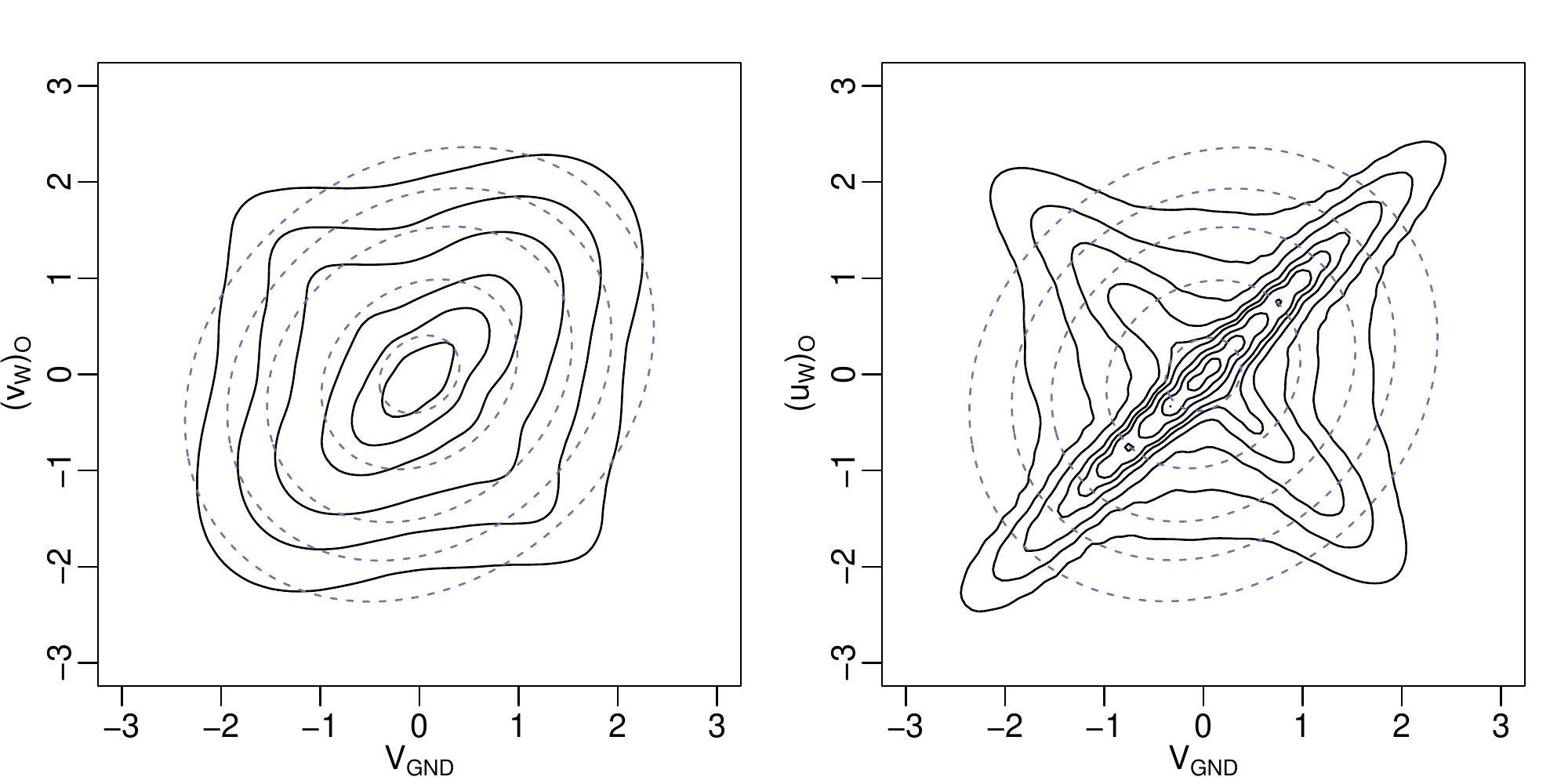}
	\caption{Copula dependence structures - ground speed and wind speeds}
    \label{ContoursSpeeds}
\end{figure}

The notable residual dependencies can be grouped into two categories. In Fig.\ \ref{ContoursAttitude}, attitude angles and associated rotational rates are given for the particular cases pitch angle \(\theta\) and pitch rate \(q\) as well as roll angle \(\phi\) and roll rate \(p\). Obviously, the attitude angle and the associated rotational rate are closely related, so the dependence among the residuals, i.e.\ measurement errors is reasonable. For the considered flights, no direct recordings of the rotational rates were given. Therefore, they were calculated based on the attitude angles in a preprocessing step. In Fig.\ \ref{ContoursSpeeds}, the ground speed \(V_{GND}\) is paired with the wind speeds \((v_W)_O\) and \((u_W)_O\) given in the NED frame \(O\). Again an obvious relation between these variables is given. Since the uncertainty of the on board recordings of the wind speeds is significant, it is not possible to completely eliminate it by the physical model, the RTS smoother framework or the methods proposed in this paper.

Both figures show kernel density contours along with contours of a reference Gaussian distribution (dashed blue). We observe two phenomena in all plots: First, the spikes in the lower left and lower right corners are slightly sharper than the Gaussian reference. This means that dependence is stronger then Gaussian when both variables show large positive or negative measurement errors simultaneously. Second, we see bumps towards the upper left and lower right corners, indicating additional dependence when there are large positive and large negative measurement errors and vice versa. This dependence is not reflected by the Gaussian distribution at all. Again, the observed shapes are characteristic for scale mixtures of Gaussians and may be captured by a more adaptive parametrization of the dynamic covariance matrix.

\section{Conclusions}\label{ConclusionChapter}
Within this paper, a detailed investigation of the measurement noise covariance of an implemented RTS smoother is carried out. The resulting noise covariance matrices are used for a reiteration of the RTS smoother. Based on a smoothing quality measure, the results could be improved in the majority of the cases. In addition, the characteristics of the residuals of the second smoother iteration are significantly closer to the assumptions of the smoother compared to the residuals of the first smoother iteration. Finally, copula models are used to characterize the remaining dependence in the residuals after the second iteration. These dependence structures are considered and potential revisions of the underlying physical and statistical models discussed.

\section*{Acknowledgements}
The research carried out in this paper was supported by the Deutsche Forschungsgemeinschaft (DFG) within the project ``Copula based dependence analysis of functional data for validation and calibration of dynamic aircraft models'' with the project identifiers CZ 86/5-1 and HO 4190/10-1.

Furthermore, the algorithms that are used as a foundation for the developments of this paper have been mainly implemented in \cite{JoachimSiegel.25.04.2017} and \cite{Siegel.18.11.2015}. Finding possibilities to characterize the measurement noise covariance matrices \(R_k\) for the application of the RTS smoother onto FDM data more precisely, which is the central goal of this paper, was mentioned in Chapter 5 ``Conclusions and Perspective'' of \cite{JoachimSiegel.25.04.2017}.

\bibliography{CopFlyBib}

\end{document}